\documentclass[a4paper]{article}
\usepackage{amsmath,epsfig}

\newcommand{\dt}{\! \cdot \!}
\newcommand{\nn}{\nonumber}
\newcommand{\bx}{\mbox{\boldmath $x$}}

\newcommand{\grad}{\nabla}
\newcommand{\bgrad}{\mbox{\boldmath $\grad$}}

\newcommand{\delsl}{\not\!\partial}
\newcommand{\gam}{\gamma}
\newcommand{\bet}{\beta}
\newcommand{\alp}{\alpha}
\newcommand{\kap}{\kappa}
\newcommand{\vareps}{\varepsilon}
\newcommand{\half}{{\textstyle \frac{1}{2}}}
\newcommand{\om}{\omega}

\newcommand{\bB}{\mbox{\boldmath $B$}}
\newcommand{\bsig}{\mbox{\boldmath $\sigma$}}
\newcommand{\bL}{\mbox{\boldmath $L$}}
\newcommand{\nr}{\mathrm{NR}}

\begin{document}

\noindent
{\bf\Large \textsf{Bound states and decay times of fermions in a
\\ Schwarzschild black hole background  }} 

\vspace{0.2cm}

\noindent 
Anthony Lasenby\footnote{e-mail: \texttt{a.n.lasenby@mrao.cam.ac.uk}}, 
Chris Doran\footnote{e-mail: \texttt{C.Doran@mrao.cam.ac.uk}}, 
Jonathan Pritchard, Alejandro Caceres \\ and Sam Dolan

\vspace{0.2cm}

\noindent
Astrophysics Group, Cavendish Laboratory, Madingley Road, \\
Cambridge CB3 0HE, UK.

\vspace{0.4cm}

\begin{center}
\begin{abstract}
We compute the spectrum of normalizable fermion bound states in a
Schwarzschild black hole background.  The eigenstates have complex
energies.  The real part of the energies, for small couplings, closely
follow a hydrogen-like spectrum.  The imaginary parts give
decay times for the various states, due to the absorption properties
of the hole, with states closer to the hole having shorter
half-lives.  As the coupling increases, the spectrum departs from that
of the hydrogen atom, as states close to the horizon become
unfavourable.  Beyond a certain coupling the $1S_{1/2}$ state is no
longer the ground state, which shifts to the $2P_{3/2}$ state, and then 
to states of successively greater angular momentum.  For
each positive energy state a negative energy counterpart exists, with
opposite sign of its real energy, and the same decay factor.  It
follows that the Dirac sea of negative energy states is decaying,
which may provide a physical contribution to Hawking radiation.
\end{abstract}

\vspace{0.2cm}

PACS numbers: 03.65.Ge, 04.70.Bw, 04.62.+v, 03.65.Pm

\end{center}

\section{Introduction}

Quantum theory in a black hole background has been extensively studied
by many authors.  Detailed discussions of this problem are contained
in the books by Birrell \& Davies~\cite{bir-quant} and
Chandrasekhar~\cite{cha83}, and the review paper by Brout \textit{et
al.}~\cite{bro95}.  Much of the attention in this work is focussed on
the wave equation and its scattering properties.  Detailed studies of
the Dirac equation in a black hole background are less common.
Indeed, the lowest order scattering cross section for a fermion in a
black hole background has only recently been
computed~\cite{DL2002,LD-erice01}.  In this paper we investigate
another previously neglected aspect of quantum mechanics in a black
hole background.  This is the existence of the bound state spectrum
for particles orbiting a spherically--symmetric point source.  That
is, we study the gravitational analogue of the hydrogen atom orbitals.

There has been strangely little effort devoted to the study of the
bound state spectrum, despite the fundamental importance of the
electromagnetic analogue.  But it is clear that these states must
exist --- how else can one provide a quantum description of a particle
in orbit around a black hole?  These states must also be essential in
the quantum description of the capture process.  The problem was
discussed in 1974 by Deruelle and Ruffini~\cite{der74}, who described
the existence of resonance states in the Klein--Gordon equation.
Further significant contributions were made in a series of papers by
Gaina and coauthors~\cite{gai87,gai88,gai92}.  These papers give
various analytic expressions for the real and imaginary parts of the
energy in a series of limiting cases.

Much of the study of quantum mechanics in a black hole background has
focussed on the related, though distinct, problem of finding the
quasi-normal mode spectrum.  Quasi-normal modes are purely ingoing at
the horizon, and outgoing at infinity~\cite{cha83}.  These boundary
conditions produce a spectrum of eigenstates with complex-valued
energies.  The significance of these quasi-normal modes comes from
their use in describing black hole oscillations.  But the boundary
condition at infinity implies that these modes are not normalizable,
so cannot represent bound states.  The problem of interest here is to
find these bound states, so we seek solutions which are purely ingoing
at the horizon, and which fall off exponentially at infinity.

For a particle of mass $m$ in the field of a black hole of mass $M$
the dimensionless coupling strength is defined by
\begin{equation}
\alp = \frac{mM}{m_p^2}
\end{equation}
where $m_p$ is the Planck mass.  In this paper we compute the fermion
bound state spectra for $\alpha$ in the range $0\cdots 6$.  If the
bound particle is assumed to be an electron, this range corresponds to
black holes of masses up to $1\times10^{15}$kg, which is the scale
appropriate for primordial black holes formed in the early universe.
Computing the energy spectrum is more complicated than the hydrogen
atom case for two main reasons.  The first is that the
radially-separated Dirac equation contains three singular points, only
two of which are regular.  There is no special function theory
appropriate for the study of such equations, so we have to resort to a
range of numerical techniques to find the spectrum.  The second
problem is that the singularity at the centre of a Schwarzschild black
hole acts as a current sink.  All normalizable states must therefore
decay in time, and we must search for eigenstates over the
two-dimensional space of complex energies.  The states we construct
therefore all have a finite half-life, so can be viewed as resonance
states.  The interpretation of these states is
discussed in section~\ref{disc}.

Despite these difficulties, the problem can be tackled numerically,
and we present a range of results for the real and imaginary parts of
the energy.  These are sufficient to predict how the spectrum will
behave for larger values of the coupling constant.  The first result,
which is entirely to be expected, is that the orbitals become
increasingly tightly bound as the coupling increases.  It follows
that, for a given state, the energy will initially decrease with
$\alp$, but will eventually turn round and start increasing as the
particle spends too much time inside the classical radius of minimum
energy.  States with higher angular momentum then become energetically
favourable as $\alp$ increases.  For example, we show that beyond
$\alp\approx 0.6$ the $1S_{1/2}$ state is no longer the ground state.
While the real part of the energy behaves in quite a complicated
fashion, the imaginary part, which controls the decay rate, simply
increases in magnitude.  This is also as one would expect.  The closer
the orbital density is to the singularity, the greater the probability
of capture.

We start by discussing the Dirac equation in a Schwarzschild
background in an arbitrary gauge.  This is helpful in establishing a
range of gauge-invariant results.  In particular, the energy conjugate
to time translation symmetry is confirmed to be a gauge invariant
quantity.  This is important in order to guarantee that the quantity
is a physical observable.  We next establish the behaviour of the
wavefunction around the horizon, which is sufficient to establish that
the states must decay exponentially with time.  We then turn to a
specific choice of gauge that is well-suited to numerical solution.
We solve the equations by simultaneously integrating out from the
horizon and in from infinity.  We then vary the energy to ensure that
the solutions match at some finite radius.  This process guarantees
that we find a global, normalizable bound state.  A set of spectra are
obtained, and the density is plotted for a range of states.  Decay
rates and expectation values for the distance are also presented.  We
end with a discussion of the significance of these bound states, and
the possible physical processes that they may generate.  Except where
stated otherwise, natural units with $G=\hbar=c=1$ are assumed
throughout.  We employ a spacetime metric with signature $-2$.

\section{The Dirac equation}

We start by defining a general parameterisation of the Schwarzschild
solution.  This general form will help to guarantee that various
expressions are gauge invariant.  We let
$\{\gam_0,\gam_1,\gam_2,\gam_3\}$ denote the standard gamma matrices
in the Dirac--Pauli representation, and introduce polar coordinates
$(r,\theta,\phi)$.  From these we define the unit polar matrices
\begin{align}
\gam_r &=  \sin\!\theta (\cos\!\phi \, \gam_1 + \sin\!\phi \,
\gam_2) + \cos\!\theta\, \gam_3  \nn \\
\gam_\theta &= \cos\!\theta (\cos\!\phi \, \gam_1 + \sin\!\phi \,
\gam_2) - \sin\!\theta\, \gam_3 \nn \\
\gam_\phi &= - \sin\!\phi \, \gam_1 + \cos\!\phi \,
\gam_2.
\end{align}
In terms of these we define the four matrices
\begin{align}
\qquad
g^t &= a_1 \gam_0 - a_2 \gam_r &
g^\theta &= - \frac{1}{r} \gam_\theta \nn \\
g^r &= -b_1 \gam_r + b_2 \gam_0 &
g^\phi &= - \frac{1}{r \sin\!\theta} \gam_{\phi} 
\qquad
\end{align}
where $(a_1,a_2,b_1,b_2)$ are scalar functions of $r$ satisfying
\begin{align}
a_1 b_1 - a_2 b_2 &= 1 \nn \\
(b_1)^2 - (b_2)^2 &= 1 - 2M/r.
\label{cnstr}
\end{align}
The reciprocal set of matrices are therefore
\begin{align}
\qquad
g_t &= b_1 \gam_0 - b_2 \gam_r &
g_\theta &= r \gam_\theta \nn \\
g_r &= a_1 \gam_r - a_2 \gam_0 &
g_\phi &= r \sin\!\theta \gam_{\phi} .
\qquad
\end{align}
These matrices satisfy
\begin{align}
\{ g^\mu, g^\nu \} &= 2 g^{\mu\nu} I \nn \\
\{ g_\mu, g_\nu \} &= 2 g_{\mu\nu} I \nn \\
\{ g^\mu, g_\nu \} &= 2 \delta^\mu_\nu I 
\end{align}
where $\mu, \nu$ run over the set $(t,r,\theta,\phi)$, $I$ is the identity
matrix, and $g_{\mu\nu}$ is the spacetime metric.  The line element
defined by this metric is
\begin{align}
g_{\mu\nu} dx^\mu dx^\nu &= 
\bigl( 1-2M/r \bigr) dt^2 + 2 \bigl(a_1 b_2-a_2 b_1\bigr) dt \,
dr - \bigl( (a_1)^2 - (a_2)^2 \bigr) dr^2 \nn \\
& \quad  - r^2( d\theta^2 + \sin^2\!\theta \, d\phi^2).
\label{Lnelm}
\end{align}
This line element is the most general form one can adopt for the
Schwarzschild solution.  There is only one degree of freedom in
equation~\eqref{Lnelm}, since the terms are related by
\begin{equation}
( 1-2M/r) \bigl( (a_1)^2 - (a_2)^2 \bigr) + (a_1 b_2-a_2 b_1)^2 = 1.
\end{equation}
This arbitrary degree of freedom corresponds to the fact that the time
coordinate is only defined up to an arbitrary radially-dependent
term.  That is, we can set
\begin{equation}
\bar{t} = t + \alpha(r),
\end{equation}
and the new line element will be independent of the new time
coordinate $\bar{t}$.  Rather than think in terms of changing the time
coordinate, however, it is simpler for our purposes to always label
the time coordinate as $t$ and instead redefine $a_1$ and $a_2$.
These then transform as
\begin{align}
a_1 \mapsto \bar{a}_1 &= a_1 - b_2 \alp' \nn \\
a_2 \mapsto \bar{a}_2 &= a_2 - b_1 \alp' ,
\label{ttrf}
\end{align}
with $b_1$ and $b_2$ unchanged.  Throughout dashes denote derivatives
with respect to $r$.  It is straightforward to confirm that the new
set $(\bar{a}_1,\bar{a}_2,b_1, b_2)$ still satisfy the constraints of
equation~\eqref{cnstr}.

The four variables $a_1$, $a_2$, $b_1$ and $b_2$ are subject to two
constraint equations, so must contain two arbitrary degrees of
freedom.  The first arises from the freedom in the time coordinate as
described in equation~\eqref{ttrf}.  The second lies in the freedom to
perform a radially-dependent boost, which transforms the variables
according to
\begin{equation}
\begin{pmatrix}
a_1 & b_1 \\
a_2 & b_2
\end{pmatrix}
\mapsto 
\begin{pmatrix}
\cosh\bet & \sinh\bet \\
\sinh \bet & \cosh\bet
\end{pmatrix}
\begin{pmatrix}
a_1 & b_1 \\
a_2 & b_2
\end{pmatrix}
\label{boost}
\end{equation} 
where $\bet$ is an arbitrary, non-singular function of $r$.  This
boost does not alter the line element of equation~\eqref{Lnelm}.  Outside
the horizon we have $|b_1| > |b_2|$, and in the asymptotically flat
region $b_1$ can be brought to $+1$ by a suitable boost.  It follows
that we must have
\begin{equation}
b_1 > 0 \quad \forall r \geq 2M.
\end{equation}
At the horizon we therefore have $b_1$ positive, and $b_2=\pm b_1$.
For black holes (as opposed to white holes) the negative sign is the
correct one, as this choice guarantees that all particles fall in
across the horizon in a finite proper time.  This sign is also
uniquely picked out by models in which the black hole is formed by a
collapse process.  We can therefore write
\begin{equation}
b_2 = - b_1 \quad \mbox{at $r=2M$}.
\end{equation}
Combining this with the identity $a_1 b_1 - a_2 b_2 = 1$ we find
that, at the horizon, we must have
\begin{equation}
a_1 b_2 - a_2 b_1 = -1 \quad \mbox{at $r=2M$}.
\label{offdg}
\end{equation}
The diagonal form of the Schwarzschild metric sets $a_1 b_2 - a_2 b_1
= 0$, so does not satisfy this criteria.  But for this case the time
coordinate $t$ is only defined outside the horizon, and the horizon
itself is not dealt with correctly.

We now have a general parameterisation of the Schwarzschild solution
in an arbitrary gauge.  The next step is to write down the Dirac
equation is this background.  This is 
\begin{equation}
i g^\mu \grad_\mu \psi = m \psi,
\end{equation}
where 
\begin{equation}
\grad_\mu \psi = (\partial_\mu + \frac{i}{2} \Gamma^{\alpha \beta}_{\mu}
\Sigma_{\alpha \beta}) \psi,
\qquad
\Sigma_{\alpha \beta} = \frac{i}{4}[\gamma_\alpha, \gamma_\beta].
\end{equation}
The components of the spin connection are found in the standard way
(see Nakahara~\cite{nak-geom}, for example).  These turn out to give
\begin{equation}
g^\mu  \frac{i}{2} \Gamma^{\alpha \beta}_{\mu}
\Sigma_{\alpha \beta} = \left( b_2' + \frac{2b_2}{r} \right) \gam_0 - 
\left( b_1' + \frac{2(b_1 -1)}{r} \right) \gam_r.
\end{equation}

For the Dirac spinor we use a radial separation of the form
\begin{equation}
\psi = \frac{e^{-iEt}}{r} 
\begin{pmatrix}
u_1(r) \chi_\kappa^\mu (\theta, \phi) \\
u_2(r) \sigma_r \chi_\kappa^\mu (\theta, \phi) 
\end{pmatrix}
\label{trispn}
\end{equation}
where $E$ is the (complex) energy and 
\begin{equation}
\sigma_r = \sin\!\theta (\cos\!\phi \, \sigma_1 + \sin\!\phi \,
\sigma_2) + \cos\!\theta\, \sigma_3.
\end{equation}
The angular eigenmodes are labeled by $\kappa$, which is a positive or
negative nonzero integer, and $\mu$, which is the total angular
momentum in the $\theta=0$ direction.  Our convention for these
eigenmodes is that
\begin{equation}
(\bsig \dt \bL + \hbar)  \chi_\kappa^\mu =  \kappa \hbar
\chi_\kappa^\mu, \quad \kappa = \ldots, -2, -1, 1,2, \ldots.
\end{equation}
The positive and negative $\kappa$ modes are related by
\begin{equation}
\sigma_r  \chi_\kappa^\mu =  \chi_{-\kappa}^\mu .
\end{equation}

The trial function~\eqref{trispn} results in the pair of coupled
first-order equations 
\begin{equation}
\begin{pmatrix} 
b_1 & b_2  \\ 
b_2 & b_1 
\end{pmatrix}
\begin{pmatrix} 
u_1' \\ u_2' 
\end{pmatrix} 
= \bB
\begin{pmatrix} 
u_1 \\ u_2 
\end{pmatrix}
\label{radeqn}
\end{equation} 
where
\begin{equation} 
\bB =
\begin{pmatrix}
\kappa/r-b_1' /2 +ia_2 E & i(m+a_1 E) -b_2'/2  \\ 
-i(m-a_1 E) -b_2'/2 & -\kappa/r-b_1'/2 +ia_2 E 
\end{pmatrix} . 
\end{equation}
These are the equations we wish to solve for complex energy $E$.  It
is first worthwhile confirming that the equations are gauge
invariant.  A redefinition of the time coordinate is equivalent to the
transformations described in equation~\eqref{ttrf}.  These are combined with
the transformation 
\begin{equation}
u_1 \mapsto u_1 e^{- iE \alp} \qquad 
u_2 \mapsto u_2 e^{- iE \alp}
\end{equation}
which together ensure that equation~\eqref{radeqn} is still satisfied.
The radial boost defined by equation~\eqref{boost} is combined with the
transformation
\begin{equation}
\begin{pmatrix}
u_1 \\
u_2
\end{pmatrix}
\mapsto
\begin{pmatrix}
\cosh(\bet/2) & -\sinh(\bet/2) \\
-\sinh(\bet/2) & \cosh(\bet/2)
\end{pmatrix}
\begin{pmatrix}
u_1 \\
u_2
\end{pmatrix}
\end{equation} 
to again ensure that the equation is still satisfied.  In either
case we see that the eigenvalue $E$ is unchanged, so is a true
gauge-invariant quantity.

The angular separation of equation~\eqref{trispn} is clearly justified
from the form of the Dirac equation.  The separation into energy
eigenstates is gauge invariant, but it is helpful to see the
separation in a gauge where the Dirac equation takes on a Hamiltonian
form.  This is provided by the `Newtonian'
gauge~\cite{DL2002,DGL98-grav} which sets
\begin{align}
\qquad a_1 &= 1 & a_2 &= 0 \nn \\
b_1 &= 1 & b_2 &= - (2M/r)^{1/2}. \qquad
\label{Newtg}
\end{align}
In this gauge the Dirac equation takes on the simple form
\begin{equation}
i \!\! \delsl \psi - i \gamma^0 \left(\frac{2M}{r} \right)^{1/2}
\left( \frac{\partial}{\partial r}  + \frac{3}{4r} \right) \psi = m \psi,
\end{equation}
where ${\not\!\partial}$ is the Dirac operator in flat Minkowski
spacetime.  This equation is manifestly separable in time, so has
solutions which go as $\exp(-iEt)$.  Since the separation works in
this gauge, it must work in all others.  We will return to this gauge
choice when we turn to finding the energy spectrum.

The nature of equation~\eqref{radeqn} can be understood more clearly by
writing it in the form
\begin{equation}
\bigl(1-2M/r \bigr)
\begin{pmatrix} 
u_1' \\ u_2' 
\end{pmatrix} 
= \begin{pmatrix} 
b_1 & -b_2  \\ 
-b_2 & b_1 
\end{pmatrix}\bB
\begin{pmatrix} 
u_1 \\ u_2 
\end{pmatrix}.
\label{radeqn2}
\end{equation}
This exposes the fact that the horizon is a regular singular point of
the radial equations.  The same is true of the origin, and infinity
turns out to be an irregular singular point.  This implies that the
radial equations cannot be manipulated into second-order
hypergeometric form, as one is able to do for the hydrogen atom.  The
closest the equations come to a recognisable form is that of Heun's
equation, which generalises the hypergeometric equation to the case of
four regular singular points on the complex plane~\cite{Ron-heun}.
But Heun's equation can usually only be analysed using numerical
techniques, and these are the tools we will apply to
equation~\eqref{radeqn}

The presence of singular points means that we must check carefully
that our solutions behave appropriately at these points.  The point at
infinity is not an issue, as we seek solutions that decay
exponentially.  Similarly, the origin is not a problem.  We expect
that the function will be weakly singular there as the origin acts as
a current sink, and this is indeed the case.  The horizon, however, is
more complicated.  The wavefunction must be well behaved at the
horizon if it is to represent a physical solution.  To test this we
introduce the series expansion
\begin{equation} 
u_1 = \eta^s \sum_{k=0}^{\infty} \alp_k \eta^k, 
\qquad u_2 = \eta^s \sum_{k=0}^{\infty} \bet_k \eta^k, 
\label{series}
\end{equation}	
where $\eta = r-2M$.  On substituting this series into
equation~\eqref{radeqn2}, and setting $\eta=0$, we obtain the indicial
equation
\begin{equation} 
\det \left[ \begin{pmatrix} b_1 & -b_2 \\ -b_2 & b_1 \end{pmatrix} \,
\bB - \frac {s}{r} I \right]_{r=2M} = 0,
\end{equation}
where $I$ is the identity matrix.  Employing the result that
\begin{equation} 
b_1 b_1' - b_2 b_2' = M/r^2
\end{equation}
we find that the two solutions of the indicial equation are 
\begin{equation} 
s=0,  -\half + 4iME (b_1 a_2-b_2 a_1)_{r=2M}.
\end{equation}
Equation~\eqref{offdg} then tells us that the two indices are
\begin{equation}
s=0,  -\half + 4iME .
\end{equation}
These indices are therefore gauge invariant.  The regular root $s=0$
ensures that we can always construct a solution that is finite and
continuous at the horizon.  The singular branch gives rise to
discontinuous solutions with an outgoing current at the horizon.
These can be used to provide a heuristic explanation of the Hawking
radiation~\cite{DGL98-grav}.  It is clear that the non-zero indicial
root gives rise to a wavefunction that is ill-defined at the horizon,
and so cannot represent a physical state.  We must therefore confine
our search for bound states to solutions that are regular at the
horizon.

The regular and singular solutions are related by a generalized form
of time-reversal symmetry.  For this we define
\begin{equation}
\bar{\psi}(t, \bx) = \frac{1}{(1-2M/r)^{1/2}} (b_1 \gamma_0 - b_2
\gamma_r) \psi^\ast(-t+f(r), \bx),
\end{equation}
which effectively reverses the time direction using the normalized
timelike Killing vector.  In terms of the $u_1$ and $u_2$ functions,
the new solution is characterised by
\begin{equation}
\begin{pmatrix} 
\bar{u}_1 \\ \bar{u}_2 
\end{pmatrix} =  \frac{\exp(-iE^\ast f(r))}{(1-2M/r)^{1/2}} 
 \begin{pmatrix} 
b_1 & b_2  \\ 
-b_2 & -b_1 
\end{pmatrix}
\begin{pmatrix} 
u_1^\ast \\ u_2^\ast 
\end{pmatrix},
\end{equation}
where $f(r)$ is determined by
\begin{equation}
(1-2M/r) \partial_r f(r) = 2(a_1 b_2 - a_2 b_1).
\end{equation}
The time-reversed solution has energy $E^\ast$ and so is exponentially
growing in time.  The solution is also is singular at the horizon,
employing the non-zero root of the indicial equation, and is not
normalizable.

Eigenmodes with different values of $\kap$ and $\mu$ are orthogonal.
For states with the same values of $\kap$ and $\mu$ the quantum inner
product can be taken as
\begin{equation} 
\langle \psi | \phi \rangle = \int_0^\infty \! dr \,   \bigl(  a_1
(u_1^\ast v_1 + u_2^\ast v_2) + a_2 (u_2^\ast v_1 + u_1^\ast v_2) \bigr),
\end{equation}
where the $u_i$ and $v_i$ denote the radial functions in $\psi$ and
$\phi$ respectively.  Current conservation for the Dirac equation is
summarised in the relation
\begin{multline}
\qquad \frac{\partial }{\partial t} \left(  a_1 (u_1 u_1^\ast + u_2
u_2^\ast) + a_2 (u_1 u_2^\ast + u_2 u_1^\ast) e^{-i(E-E^\ast)t} \right) \\ 
= - \frac{\partial }{\partial r} \left( b_1 (u_1 u_2^\ast + u_2
u_1^\ast) + b_2 (u_1 u_1^\ast + u_2 u_2^\ast) e^{-i(E-E^\ast)t}
\right). \quad 
\end{multline}
Again it is straightforward to confirm that this equation is gauge
invariant.  The right-hand side of this equation defines 
$r^2$ times the radial flux.  We denote this by $J$,
\begin{equation}
J(r) = b_1 (u_1 u_2^\ast + u_2 u_1^\ast) + b_2 (u_1 u_1^\ast + u_2
u_2^\ast).
\end{equation}
For spatially normalizable states we must have $J\mapsto 0$ as $r
\mapsto \infty$.  But at the horizon we also have
\begin{equation}
J = - b_1 | u_1 - u_2 |^2 ,
\end{equation}
which defines an inward-pointing current.  At the horizon, the regular
solution has
\begin{equation}
| u_1 - u_2 |^2 = | \alp_0 - \bet_0|^2,
\end{equation}
using the power series expansion of equation~\eqref{series}.  The
coefficients are related by
\begin{equation}
\left( iE - \frac{1}{8M} + b_1 \Bigl( \frac{\kappa}{2M} -im \Bigr)
\right) \alp_0 = \left( - iE + \frac{1}{8M} + b_1 \Bigl(\frac{\kappa}{2M}
-im \Bigr) \right) \bet_0 .
\end{equation}
It is therefore impossible to satisfy $\alp_0 = \bet_0$ for finite
energy, so there must be a non-vanishing inward current present at the
horizon.  This in turn tells us that the state must decay.  This decay
takes place at the origin, where unitary evolution breaks
down~\cite{DGL98-grav,san77}.  For bound states the energy $E$ must
contain real and imaginary terms, so we set
\begin{equation}
E = \om -i \nu.
\end{equation}
Current conservation now takes the form
\begin{equation}
 \frac{d J }{d r}
= 2 \nu \bigl(  a_1 (u_1 u_1^\ast + u_2 u_2^\ast)
+ a_2 (u_1 u_2^\ast + u_2 u_1^\ast) \bigr).
\end{equation}

Given a set $(u_1,u_2,E,\kappa)$ which solve the radial
equation~\eqref{radeqn} a new solution set is generated by the
transformation 
\begin{equation}
(u_1,u_2,E,\kappa) \mapsto (u_2^\ast,u_1^\ast,-E^\ast,-\kappa).
\end{equation}
It follows that the real part of the energy spectrum is symmetric
about the zero point.  That is, for a state with real energy $\om$ a
corresponding antiparticle state exists with real energy $-\om$.  The
decay rate is the same for both states, however.  If we assume that
the vacuum is constructed from the Dirac sea of negative energy
states, then this vacuum will decay in time.  A loss of negative
energy states can be equally interpreted as generation of positive
energy states, which provides a suggestive physical model for Hawking
radiation.

\section{The energy spectrum}

To solve for the energy spectrum we work mainly in the Newtonian gauge
of equation~\eqref{Newtg}.  In this gauge the interaction with the black
hole is defined solely by an interaction Hamiltonian
$H_{\mbox{\scriptsize I}}$ given by
\begin{equation}
H_{\mbox{\scriptsize I}} \psi = i \hbar \left( \frac{2GM}{r}
\right)^{1/2} \frac{1}{r^{3/4}} \frac{\partial}{\partial r} \bigl(
r^{3/4}  \psi \bigr). 
\end{equation}
Dimensional constants are included in a number of equations in this
section to illustrate certain features of the problem.  The line
element for the Newtonian gauge has flat spatial sections for constant
$t$, so the quantum inner product between states has the simple
flat-space form
\begin{equation}
\langle \psi | \phi \rangle = \int \! d^3x \, \psi^\dagger \phi .
\end{equation}
The interaction Hamiltonian is not Hermitian, as we have
\begin{equation}
H_{\mbox{\scriptsize I}} - H^\dagger_{\mbox{\scriptsize I}} = 
-i \hbar (2GMr^3)^{1/2} \delta(\bx).
\end{equation}
It is straightforward to check that all wavefunctions approach the
origin as $r^{-3/4}$, so the non-Hermitian part of
$H_{\mbox{\scriptsize I}}$ has finite expectation.  This confirms that
Hermiticity only breaks down at the origin, as stated earlier.  In
this respect it may be more natural to refer to $H_{\mbox{\scriptsize
I}}$ as a `pseudo-Hamiltonian', in the sense of an operator acting on
an open quantum system~\cite{exn-open}.  There is no doubt that the
system described here is open, as the singularity is not treated as
part of the quantum system.  But the system is only open in an
extremely simple fashion.  There is no ambiguity in either the time
evolution of the state, or the correct definition of the observable
energy.  Time evolution is defined by the Dirac equation, in whichever
gauge one chooses to adopt, and the energy is defined by the
energy-momentum tensor.  The fact that we have a Hamiltonian
description at all is a result of a series of gauge choices, so one
must be careful not to place too strong an interpretation on the
gauge-dependent quantity $H_{\mbox{\scriptsize I}}$.

The bound state energy eigenspectrum is determined entirely by the
properties of the wavefunction at the horizon and at infinity.  The
demands that the wavefunction is finite at the horizon and falls off
exponentially at infinity are sufficient to produce the spectrum.  But
it is only by considering the global properties of the wavefunction
that the imaginary contribution to the energy is fully understood.
Decay only takes place at the singularity, and the decay rate for a
given eigenstate is naturally related to the behaviour of the
wavefunction near the singularity.  If the spatial degrees of freedom
in an energy eigenstate are normalized such that
\begin{equation}
\int \! dr \, (u_1 u_1^\ast + u_2
u_2^\ast) = 1
\end{equation}
then the imaginary component of the energy, $-i\nu$, is determined by
\begin{equation}
\nu = \lim_{r \mapsto 0} \frac{\hbar(2GM)^{1/2}}{2} \frac{1}{r^{3/2}}
(u_1 u_1^\ast + u_2 u_2^\ast).
\label{nufrmnorm}
\end{equation}
This identity only holds if the state is globally normalized.  It
provides a further independent check that the solutions we obtain
numerically are globally normalizable bound states.  The decay rate
increases for states with a larger probability density near the
singularity, as one would expect.  The fact that the states approach
the origin as $r^{-3/4}$ ensures that the radial probability density
tends smoothly to zero at the singularity.  The presence of the
singularity does not prevent the formation of normalizable states, and
the singular nature of the wavefunction is no worse than that of the
ground state of the hydrogen atom.

The interaction Hamiltonian is independent of the speed
of light, so the non-relativistic approximation to the Dirac equation
results in the Schr\"{o}dinger equation
\begin{equation}
  -\frac{\hbar^2 \bgrad^2}{2m} \psi_{\nr} + i \hbar \left(
\frac{2GM}{r} \right)^{1/2} \frac{1}{r^{3/4}} \frac{\partial}{\partial
r} \bigl( r^{3/4} \psi_{\nr} \bigr) = E_{\nr} \psi_{\nr} ,
\end{equation}
where the subscript NR denotes non-relativistic.  If we now introduce
the phase-transformed variable
\begin{equation} 
\Psi=\psi_{\nr}\exp \left(- i(8r/a_0)^{1/2}\right) 
\end{equation}  
where
\begin{equation}
a_{0}=\frac{\hbar ^{2}}{GMm^{2}}
\end{equation}
we see that $\Psi$ satisfies 
\begin{equation}
-\frac{\hbar^2 \bgrad^2}{2m} \Psi - \frac{GMm}{r} \Psi = E_{\nr} \Psi.
\end{equation} 
In the non-relativistic limit the energy spectrum is therefore given
by the gravitational analogue of the hydrogen atom spectrum~\cite{gai88},
\begin{equation}
E_{\nr} = - \frac{G^2 M^2 m^3}{2 \hbar^2} \frac{1}{n^2}, \quad n
= 1, 2, \ldots.
\end{equation}
In terms of the Planck mass $m_p$ we can also write
\begin{equation}
E_{\nr} = - \left( \frac{Mm}{m_p^2} \right)^2 \frac{mc^2}{2n^2}.
\label{Enr}
\end{equation}
The fact that we have a reasonable starting point for the spectrum in
the weak-coupling limit is valuable, as our method involves searching
for eigenvalues over the complex energy plane.  By analogy with the
hydrogen atom case, we expect that the non-relativistic spectrum will
be a reasonable approximation provided
\begin{equation}
 \frac{Mm}{m_p^2} \ll 1.
\end{equation}

Returning to the full, relativistic equation~\eqref{radeqn}, we 
convert this to dimensionless form by introducing the dimensionless
distance variable
\begin{equation}
x = \frac{r c^2}{GM},
\end{equation}
which ensures that the horizon lies at $x=2$.  We also introduce the
dimensionless coupling coefficient
\begin{equation}
\alp = \frac{Mm}{m_p^2}
\end{equation}
and energy
\begin{equation}
\vareps = \frac{EM}{c^2 m_p^2}.
\end{equation}
In terms of these our eigenvalue problem becomes
\begin{align}
(x-2) \begin{pmatrix} 
u_1' \\ u_2' 
\end{pmatrix} 
&= 
\begin{pmatrix}
1 & (2/x)^{1/2} \\
(2/x)^{1/2} & 1
\end{pmatrix}
\nn \\
& \quad \cdot
\begin{pmatrix}
\kappa & ix(\alp+\vareps) - (8x)^{-1/2}  \\ 
-ix(\alp-\vareps)- (8x)^{-1/2} & -\kappa
\end{pmatrix} 
\begin{pmatrix} 
u_1 \\ u_2 
\end{pmatrix}
\label{dmleqn}
\end{align} 
where the dashes now denote derivatives with respect to $x$.  We seek
eigenvalues $\vareps$ for fixed coupling $\alp$.

Two complementary methods are employed to solve the eigenvalue
problem.  We start with a series expansion around the horizon of the
regular branch of the solution.  The restriction to this branch
removes two degrees of freedom at the horizon, so the function is
uniquely specified up to an overall magnitude and phase.  These are
chosen conveniently by setting $u_1=1$ at the horizon.  The power
series expansion extends the solution a short distance away from the
horizon, from where the values of $(u_1,u_2)$ can be used to initiate
numerical integration of the differential equation~\eqref{dmleqn}.  For
most values of $\vareps$ the numerical integrator will start to
increase exponentially after a finite distance.  The aim initially is
to vary $\vareps$ so as to push this distance out as far as possible.
This requires a reasonable initial guess for the eigenvalues, which is
where the non-relativistic approximation is helpful to get things
started.

Once we have achieved a reasonably accurate value for $\vareps$, we
turn to a more sophisticated method to improve accuracy.  We seek
normalizable states for which $\psi$ is finite over all space.  To be
confident we have found such a state we need to numerically integrate
inwards from infinity, as well as outwards from the horizon.  If the
solutions for $u_1$ and $u_2$ can be arranged to match at some
suitable radius then we have found a global solution to the
first-order equations~\eqref{dmleqn}.  To expand about infinity we need to
take care of the essential singularity present there.  A suitable
series expansion is provided by
\begin{equation}
\begin{pmatrix}
u_1 \\ u_2
\end{pmatrix}
= \exp\Bigl(-px + 2i\vareps (2x)^{1/2} + \frac{\alp^2-2p^2}{p} \ln x \Bigr)
\sum_{n=0}
\begin{pmatrix}
\alp_{n/2} x^{-n/2} \\
\beta_{n/2} x^{-n/2} 
\end{pmatrix}
\label{infser}
\end{equation}
where
\begin{equation}
p^2 = \alp^2 - \vareps^2 = \frac{M^2}{m_p^4 c^4}(m^2c^4 - E^2).
\end{equation}
The definition of $p$ involves a complex square root, and the branch
is chosen so that $p$ has a positive real value, ensuring the
wavefunction falls off exponentially.

The fact that only one root of the indicial equation is used implies
that, for a given $\vareps$, $\psi$ is specified at infinity up to an
arbitrary magnitude and phase.  The first few terms in the series
expansion~\eqref{infser} are used to compute $\psi$ at a finite radius and
these values are then numerically integrated inwards.  A certain
amount of fine tuning is then required to pick the radius at which to
attempt matching.  Once a radius is chosen the matching condition is
that the inward and outward values of the two complex functions $u_1$
and $u_2$ agree.  This condition is converted into a set of four
scalar equations which state that the real and imaginary differences
vanish.  In addition we have four arbitrary parameters to vary --- the
real and imaginary terms in the energy, and the magnitude and phase of
the function integrated inwards from infinity.  This system of four
equations and four unknowns is then solved by a Newton--Raphson
method.  This converges very quickly and affords good control over
accuracy.

Three independent checks were performed on the energy spectrum achieved
by this method.  The first was that the calculations were repeated
using the same scheme in a different gauge.  The gauge chosen for
comparison is defined by advanced Eddington--Finkelstein coordinates,
with
\begin{align}
\qquad a_1 &= 1+M/r & a_2 &= M/r \nn \\
b_1 &= 1-M/r & b_2 &= - M/r. \qquad
\end{align}
The second test involved using a minimax routine to find the energy
spectrum.  This method is less accurate, but gave good agreement for
the states of lowest energy.   The final check was to confirm that,
after normalization, the states satisfy the identity of
equation~\eqref{nufrmnorm}.  This check was again satisfied to high
precision.

\section{Results}

\begin{figure}
\begin{center}
\includegraphics[height=12cm]{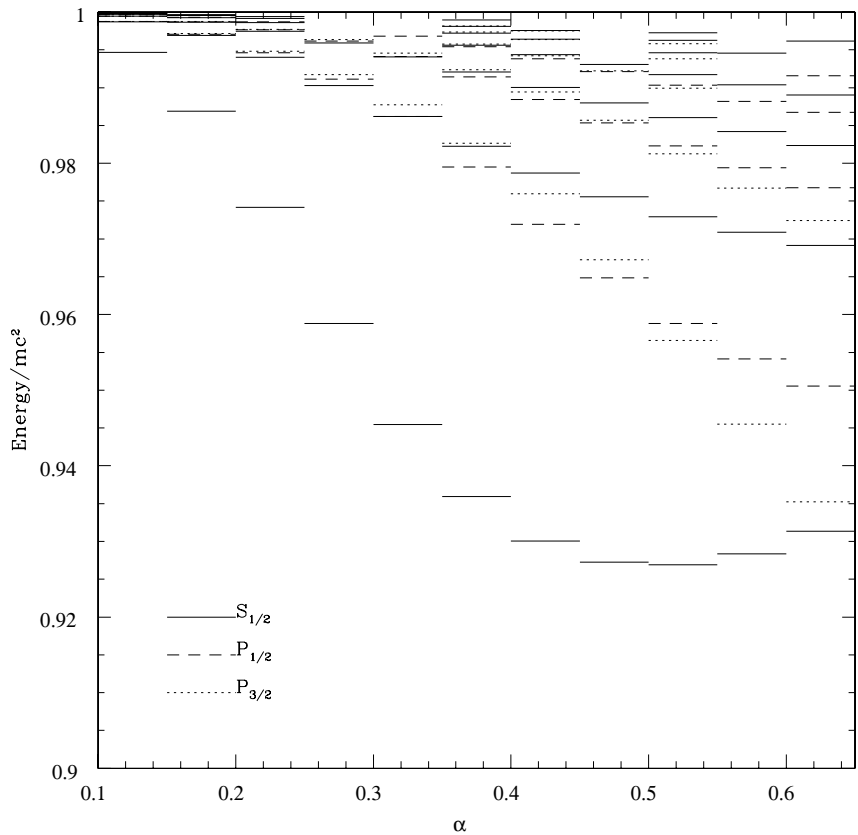}
\end{center}
\caption[dummy1]{The real part of the bound state energy, in units of
$mc^2$.  The horizontal axis labels the dimensionless coupling
coefficient $\alp$, and the lines represent the value of the energy
for the coupling at the left of the line, with $\alp$ ranging from
$0.1$ to $0.6$ in steps of $0.05$.  The $S_{1/2}$, $P_{1/2}$ and
$P_{3/2}$ orbits are shown.}
\label{Fig1}
\end{figure}

The real parts of the energy for the three lowest-energy states are
plotted in figure~\ref{Fig1}.  The vertical axis plots the real part
of the energy in units of the rest energy of the particle, which is
given by
\begin{equation}
\frac{E}{mc^2} = \frac{\vareps}{\alp}.
\end{equation}
The fact that we obtain this dimensionless ratio reflects the
equivalence principle.  The mass $m$ does not effect the spectrum on
its own --- the spectrum only depends on the product $mM$.  States are
labelled using the standard spectroscopic scheme.  In this scheme
$\kap=1$ corresponds to $S_{1/2}$, $\kappa=2$ to $P_{3/2}$ and
$\kappa=-1$ to $P_{1/2}$.  For each eigenvalue $\kap$ a ladder of
levels is obtained.

\begin{figure}
\begin{center}
\includegraphics[width=12cm]{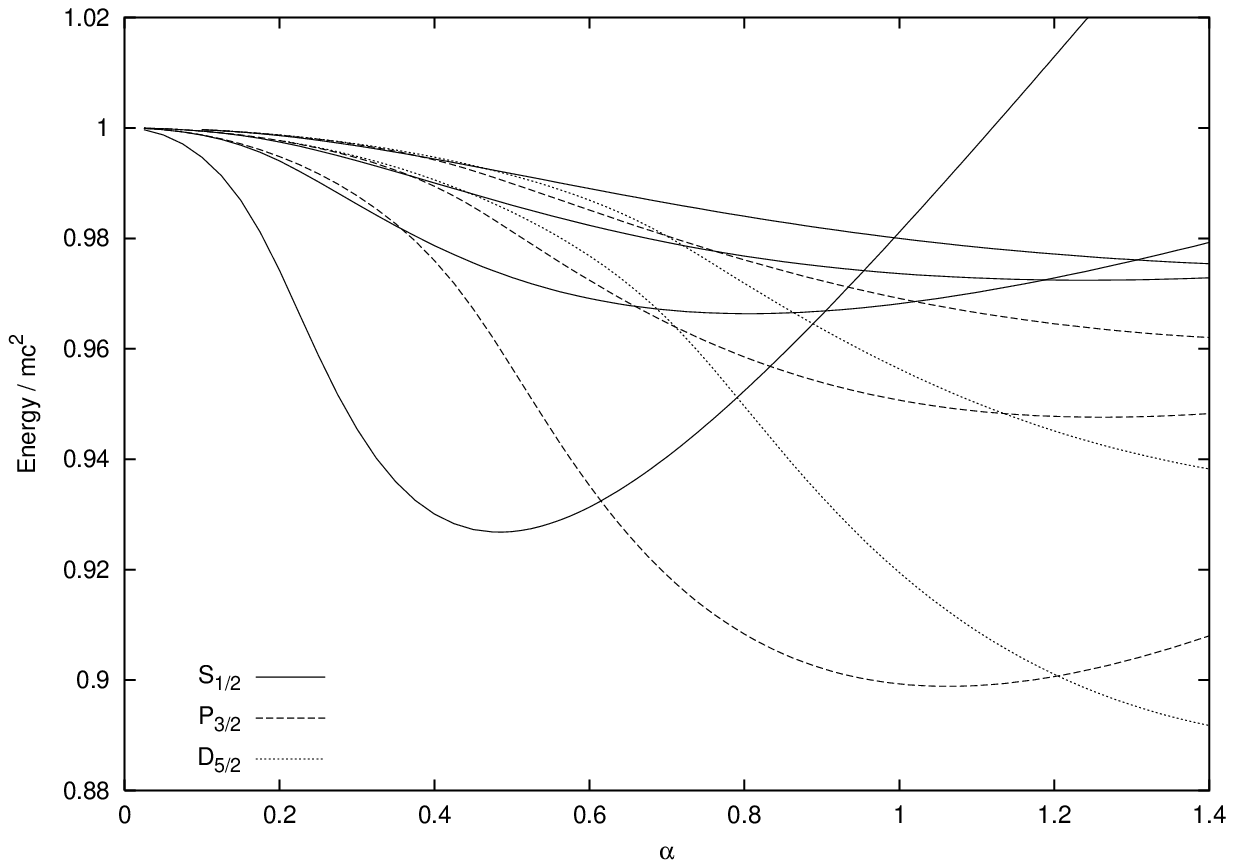}
\end{center}
\caption[dummy1]{The energy spectra of the $S_{1/2}$, $P_{3/2}$ and
$D_{5/2}$ states. At around $\alp = 0.6$, the $2P$ state becomes the 
ground state, and beyond $\alp = 1.2$ it is replaced by the $3D$ state.}
\label{Fig2}
\end{figure}

The energy spectrum illustrates a number of remarkable features.  For
small $\alp$ the spectrum resembles that of a hydrogen atom.  But as
the coupling increases the energy of the $1S_{1/2}$ state reaches a
minimum and then starts to increase.  The gravitational case avoids
the $Z=137$ catastrophe of the relativistic Coulomb problem.  This is
to be expected --- coupling strengths with $\alp > 1$ are routinely
achieved astrophysically and such objects appear to be stable.  We
also see that as $\alp$ increases beyond $0.6$ the $P_{3/2}$ state
appears to take over as the ground state.  This is confirmed in
figure~\ref{Fig2}, which shows the spectra of the $S_{1/2}$, 
$P_{3/2}$ and $D_{5/2}$ states out to $\alpha = 1.4$.  We see clearly that around $\alp=0.6$ the
$2P_{3/2}$ state takes over from $1S_{1/2}$ as the ground state, only 
to be replaced in turn by the $3D_{5/2}$ state at $\alp=1.2$.  An 
explanation of this phenomena can be found in the classical expression 
for the binding energy in a Schwarzschild potential.

\begin{figure}
\begin{center}
\includegraphics[width=6cm]{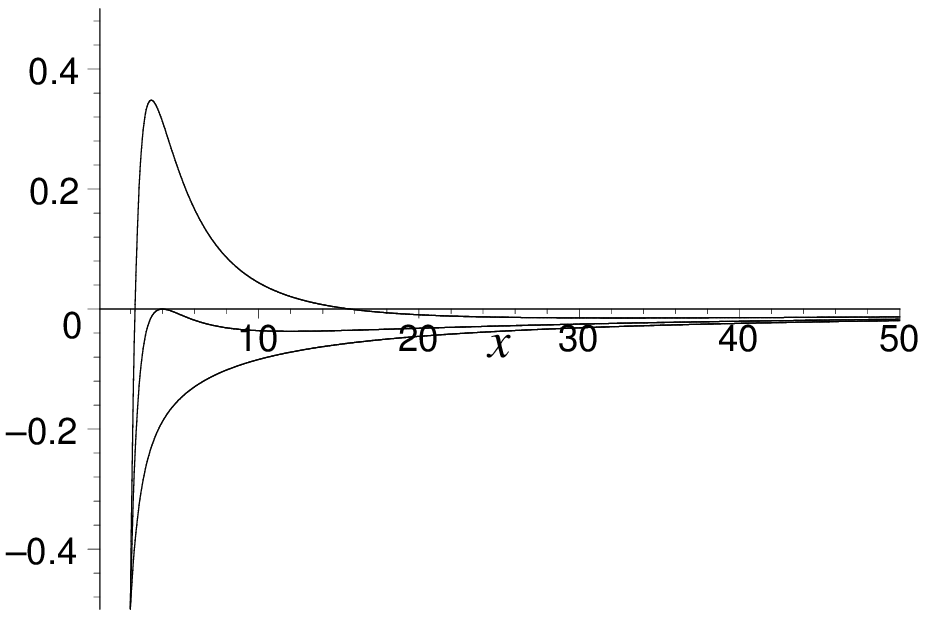}
\includegraphics[width=6cm]{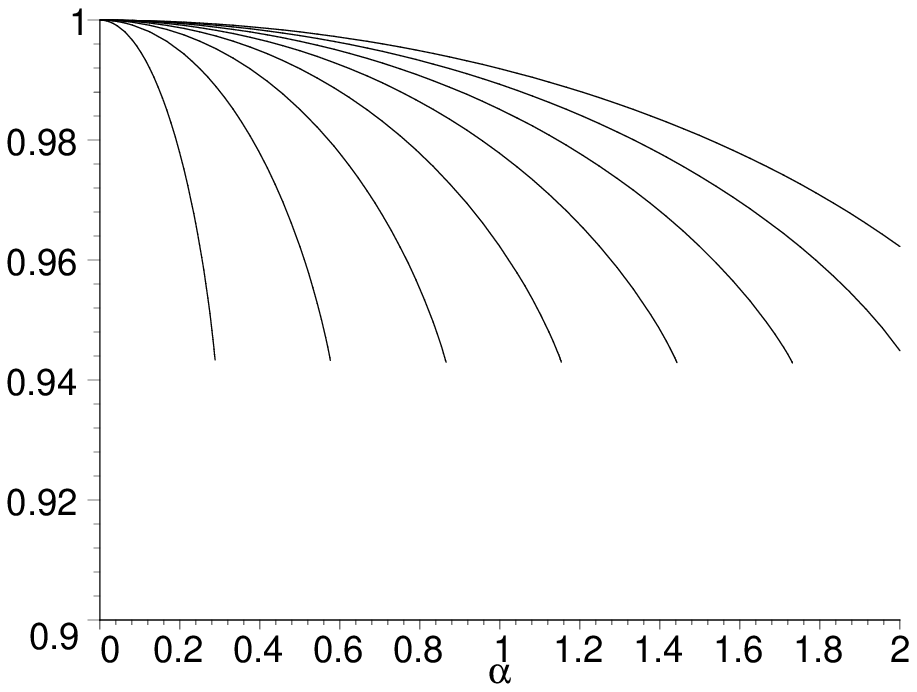}
\end{center}
\caption[dummy1]{Bohr quantisation of classical circular
orbits. The left hand plot shows the effective potential in units of
$mc^2$ for $\alpha=0.5$ and $J=n \hbar$, with $n=1,2,3$.  For $n=1$
no classical bound state is possible.  As $n$ is increased a minimum
forms in the potential, and the barrier between the minimum and the
singularity grows larger.  The right-hand plot shows the relativistic
energy in units of $mc^2$ as a function of $\alpha$.  The first eight
$n$ values are shown.  For each $n$ value the minimum energy is
achieved when $\alpha=n/\sqrt{12}$.}
\label{fVeff}
\end{figure}

For a particle of mass $m$ in a Schwarzschild background the dynamics
reduces to motion in the effective radial potential
\begin{equation}
V_{\mathit{eff}} = -\frac{GMm}{r} + \frac{J^2}{2mr^2}
\left(1-\frac{2GM}{c^2 r} \right),
\label{Egrv-veff}
\end{equation}
where $J$ is the angular momentum of the particle.  This is
illustrated in figure~\ref{fVeff}.  For $J>\sqrt{12}GMm/c$, classical
bound states can exist as the effective potential has a minimum, but
if the particle's angular momentum is smaller than $\sqrt{12}GMm/c$ it
becomes insufficient to support a classical orbit.  For a circular
orbit at radius $r$ the conserved relativistic energy, conjugate to
time translation, is
\begin{equation}
E = mc^2 \frac{r-2GM/c^2}{r^{1/2} (r-3GM/c^2)^{1/2}}.
\end{equation}
The radius $r$ and angular momentum $J$ are related by
\begin{equation}
\frac{J^2}{m^2} = \frac{GMr^2}{r-3GM/c^2}.
\end{equation}

Now suppose we attempt a form of naive Bohr quantisation by setting
\begin{equation}
J = n \hbar.
\end{equation}
Converting to dimensionless quantities the effective potential
becomes
\begin{equation}
\frac{V_{\mathit{eff}}}{mc^2} = -\frac{1}{x} + \frac{n^2}{2\alpha^2 x^2}
\left(1-\frac{2}{x} \right),
\end{equation}
and the orbital energy is
\begin{equation}
\frac{\vareps}{\alp} = \frac{x-2}{\bigl(x(x-3)\bigr)^{1/2}}
\end{equation}
where 
\begin{equation}
x = \frac{n^2}{2 \alp^2} \left( 1+ \Bigl( 1 - \frac{12
\alp^2}{n^2} \Bigr)^{1/2} \right).
\end{equation}
In the small $\alpha$ regime this reproduces the spectrum of
equation~\eqref{Enr}.  But as $\alp$ increases the energy falls to a
minimum at $\alp^2=n^2/12$, beyond which the orbit no longer exists
for a given $n$ (see figure~\ref{fVeff}).  The minimum energy achieved
is $0.94mc^2$, corresponding to $x=6$.  Inside this radius no stable
classical circular orbits exist.  In the quantum description we find
that as $\alp$ increases the orbits get more tightly bound around the
horizon.  As the coupling increases the orbits are dominated by terms
inside $x=6$ and so become energetically less favourable.  The ground
state is then one of higher angular momentum, for which the orbit is
less tightly bound.  Figure~\ref{Fig2} also shows that as $\alpha$
increases the $1S_{1/2}$ state becomes unbound.  This effect is also seen
classically, as circular orbits with $r<4M$ are known to be unbound,
as well as unstable.

The form of the effective potential illustrates a further feature of
the quantum states, which is that the quantum decay can be interpreted
as a tunnelling phenomena.  This is certainly a valid picture for
states with $n > \sqrt{12}\alpha$.  For a fixed $n$, as $\alpha$
increases, the potential barrier decreases and we expect that the
tunneling rate onto the singularity will increase.  This is indeed the
case, as we discuss further in section~\ref{decay}.

\begin{figure}
\begin{center}
\includegraphics[width=10cm]{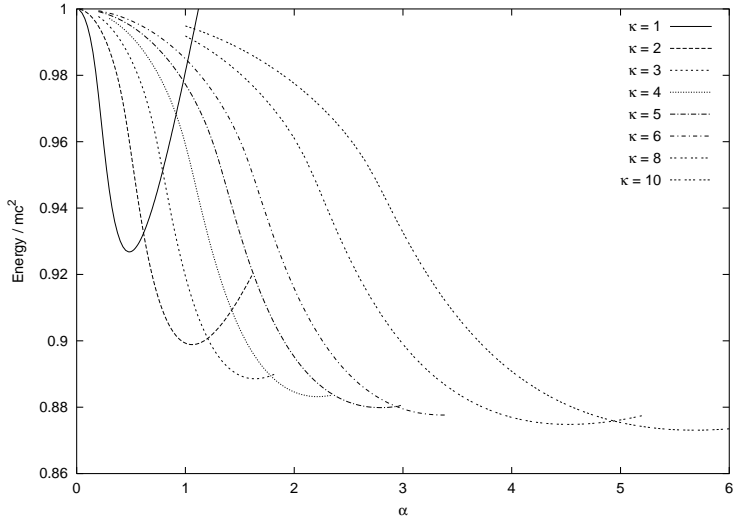}
\end{center}
\caption[dummy1]{\textit{Energy levels of states with higher angular momenta}. This plot shows the energy levels of the lowest energy states with a range of angular momenta $\kappa = 1\cdots10$. It illustrates how each state takes a turn as the groundstate, as the coupling is increased. The positions of the energy minima are linearly-spaced in $\alpha$ and have increasingly large binding energies.}
\label{higher_kappa}
\end{figure}

Figure~\ref{higher_kappa} shows how states of successively higher
angular momentum take over as the ground state as the coupling is
increased. In the small $\alpha / \kappa$ limit, the energy levels
resemble those of the classical orbits. At larger couplings, the
energy of a given state falls to a minimum, and then begins to
increase again, apparently without limit.  The $\alpha$ value at which
the minimum occurs is roughly proportional to the angular momentum of
the state, with $\alpha = 0.58 \kappa - 0.10$ providing a good
fit. The maximum binding energy available increases with angular
momentum, to beyond $E=0.88mc^2$.  This means that quantum mechanics
predicts around twice the classical value for the radiation efficiency
of accretion processes.  To confirm this effect we need to find the
limiting value of the binding energy for astrophysical values of
$\alpha$.  For an electron around a solar-mass black hole, for
example, we have $\alpha = 4 \times 10^{15}$, so a large $\alpha$
limit of our equations should be very accurate.

\section{Wavefunction properties}

\begin{figure}
\begin{center}
\includegraphics[height=11cm]{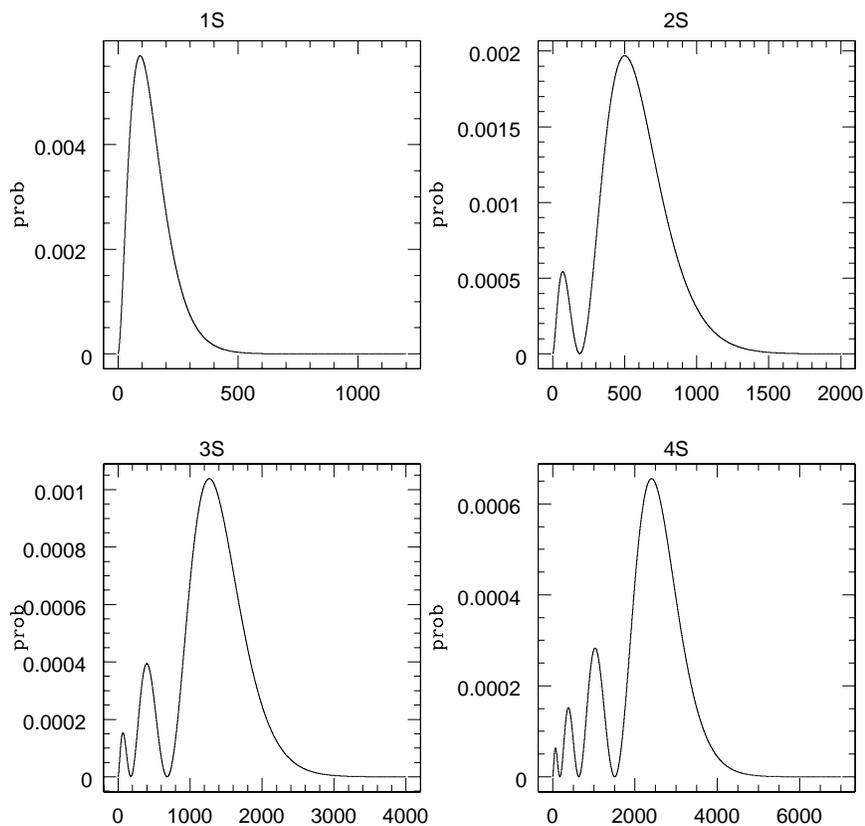}
\end{center}
\caption[dummy1]{The radial probability density for the $1S_{1/2}$,
$2S_{1/2}$, $3S_{1/2}$ and $4S_{1/2}$ states for a coupling of
$\alp=0.1$ .  The horizontal axis is the dimensionless radius $x$, and
the horizon lies at $x=2$.  All plots are started from the horizon.
The part of the density inside the horizon is not plotted, though in
all cases this smoothly approaches the origin.}
\label{FRadp1}
\end{figure}

With our current choice of gauge the radial form of the wavefunction
is best visualised by plotting $r^2$ times the timelike component of
the current.  We denote this $\rho$, so
\begin{equation}
\rho = |u_1|^2+|u_2|^2.
\end{equation}
The gauge invariant definition of $\rho$ is that it is $r^2$ times the
density as measured by observers in radial free-fall from rest at
infinity.  The first four $S_{1/2}$ states for small coupling are
shown in figure~\ref{FRadp1}.  The plots are very similar to those for
the non-relativistic hydrogen atom.  In all cases the peak of the
wavefunction is a long way outside the horizon, with only a small
fraction of the probability density lying inside the horizon.

\begin{figure}
\begin{center}
\includegraphics[height=11cm]{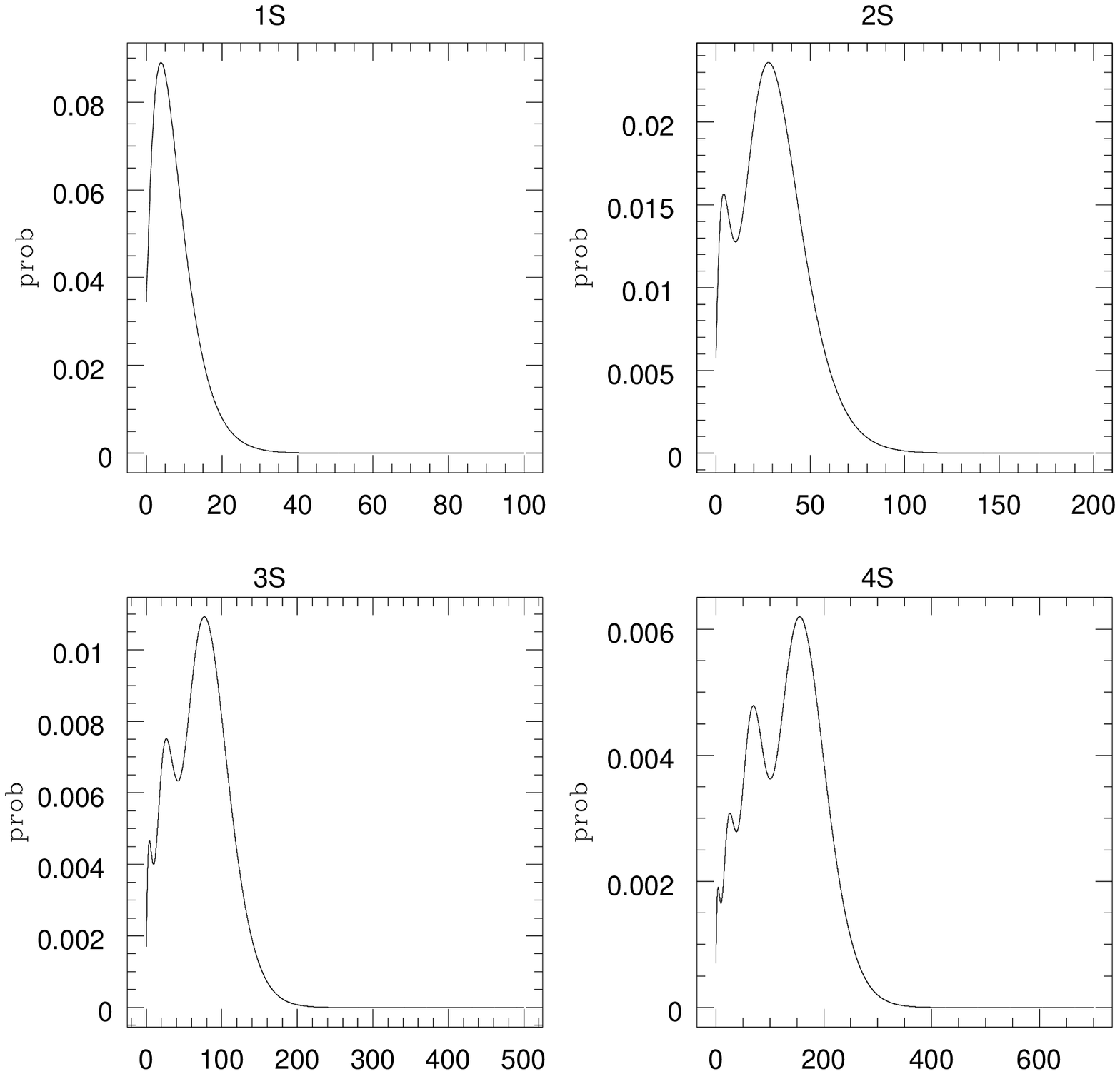}
\end{center}
\caption[dummy1]{The radial probability density for the $1S_{1/2}$,
$2S_{1/2}$, $3S_{1/2}$ and $4S_{1/2}$ states for a coupling of
$\alp=0.35$.  The nodal pattern seen in figure~\ref{FRadp1} is
beginning to get washed out as the wavefunction compresses around the
horizon.}
\label{FRadp2}
\end{figure}

\begin{figure}
\begin{center}
\includegraphics[height=11cm]{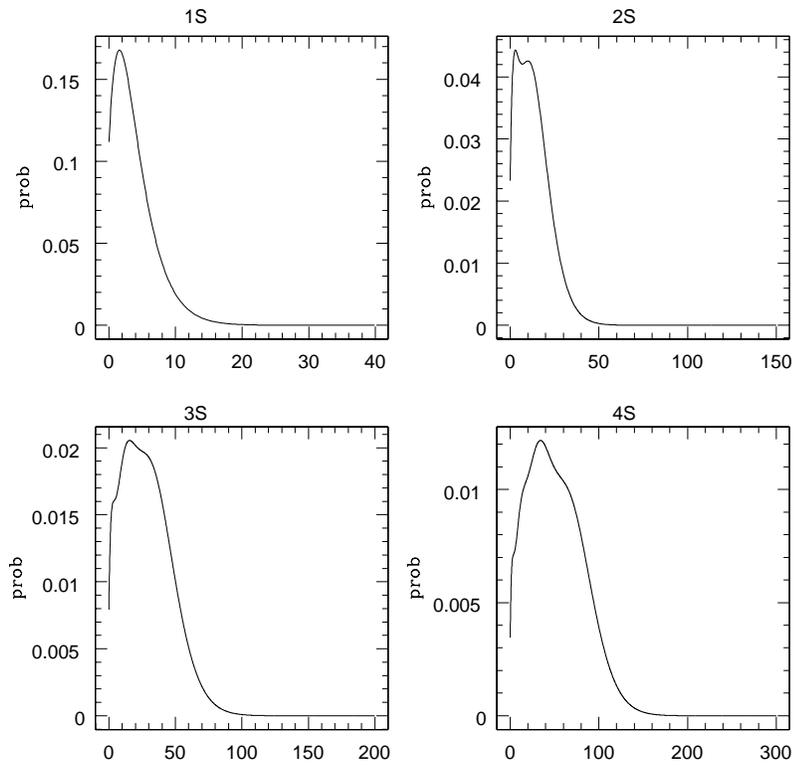}
\end{center}
\caption[dummy1]{The radial probability density for the $1S_{1/2}$,
$2S_{1/2}$, $3S_{1/2}$ and $4S_{1/2}$ states for a coupling of
$\alp=0.5$.  The pattern of nodes and dips seen in
figures~\ref{FRadp1} and~\ref{FRadp2} has almost completely vanished,
leaving a series of density profiles that lack structure.}
\label{FRadp3}
\end{figure}

As we increase the coupling to $\alp=0.35$ we obtain the series of
plots in figure~\ref{FRadp2}.  Predictably, the wavefunctions start to
bunch in closer to the horizon.  Slightly more surprisingly, the nodal
structure disappears for larger couplings.  The density
no longer drops down to near zero at a number of nodes, but instead a
number of dips are present.  If we increase the coupling further
still, to $\alp=0.5$, the dips themselves are largely washed out and
we obtain the somewhat structure-less plots shown in
figure~\ref{FRadp3}.

\begin{figure}
\begin{center}
\includegraphics[height=8cm]{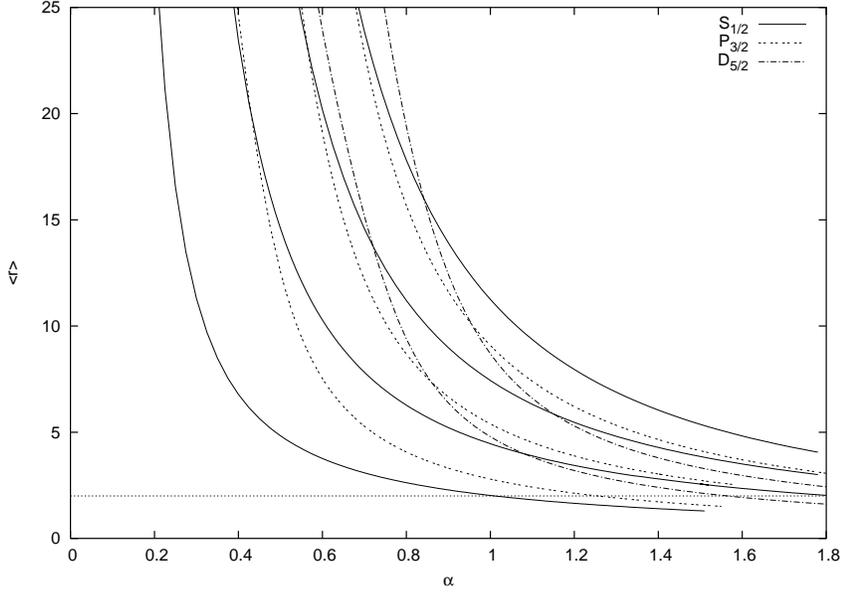}
\end{center}
\caption[dummy1]{The expectation value of $r$ in units of $GM/c^2$
for the $S$, $P$ and $D$ states.  The broken
horizontal line at x=2 shows the position of the horizon.}
\label{Fexpr}
\end{figure}

Some additional insight into the nature of the orbitals is
obtained by calculating the expectation value of $r$.  With our
current gauge choices this is defined in the obvious manner as
\begin{equation}
\langle r \rangle = \frac{\int_0^\infty \! dr \, r(|u_1|^2 +
|u_2|^2)}{\int_0^\infty \! dr \, (|u_1|^2 + |u_2|^2)}.
\end{equation}
These are calculated via a straightforward Simpson's rule, and the
results for the $S$, $P$ and $D$ orbitals are shown in
figure~\ref{Fexpr}.  We see that $\langle r \rangle$ decreases as the
coupling increases. In the low-alpha regime, the expectation value
follows the radius of the classical circular orbit, so $\langle r
\rangle \propto \alpha^{-2}$. As the coupling increases, and stable
orbits become classically impossible, we find that $\langle r \rangle$
approaches, and moves within, the horizon.  For higher $\alpha$ the
bulk of the probability density lies inside the horizon,
representing a short-lived state of a tightly-bound particle.

\begin{figure}
\begin{center}
\includegraphics[height=11cm]{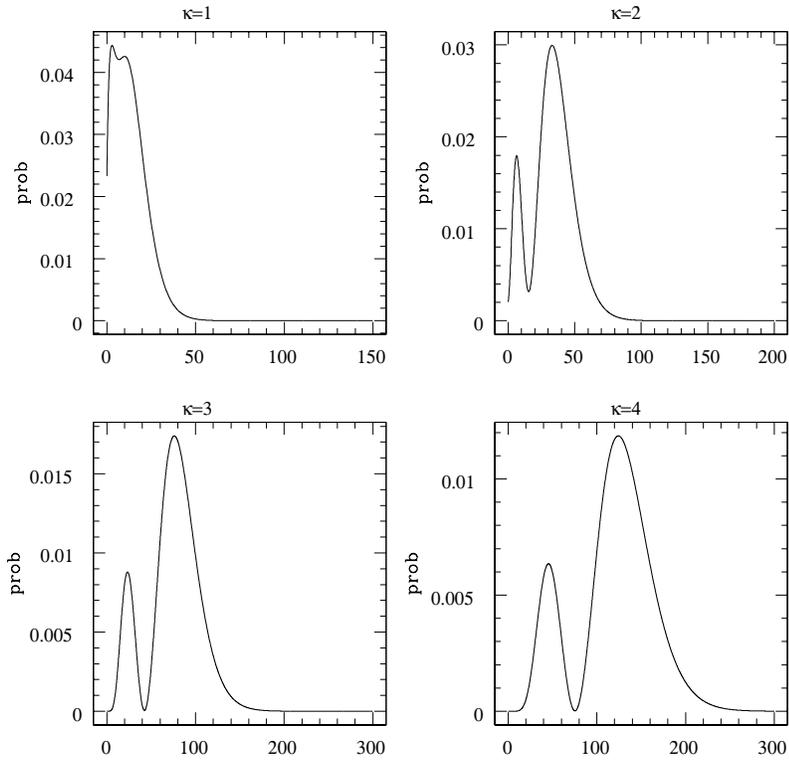}
\end{center}
\caption[dummy1]{The radial probability density for a range of angular
momentum values with a coupling of $\alp=0.5$.  The first excited
states are shown for $\kappa=1,2,3,4$.  As $\kap$ increases the
orbitals are concentrated further from the source, and begin to
resemble hydrogen atom wavefunctions.}
\label{Fradp4}
\end{figure}

While the low angular momentum orbitals are concentrated near the
horizon, the orbitals with larger angular momentum still lie an
appreciable distance out.  As such, they adopt a form closer to the
familiar hydrogen atom orbitals.  A series of such orbitals are shown
in figure~\ref{Fradp4}, which shows the first excited mode for $\kappa$
values of 1, 2, 3 and 4.  The coupling is again set to $0.5$.  As
expected, the probability density is concentrated successively further
from the hole.  By the time we reach $\kap=3$ (a classical radius of
$x=33$) the wavefunction returns to the familiar hydrogen-like form.

\section{Decay rates}
\label{decay}

So far we have concentrated on the real part of the energy, and the
associated orbitals.  But the fact that the black hole effective
Hamiltonian is not Hermitian implies that the energy is not real and
the states have a finite half-life.  As such the solutions could be
viewed as representing resonance states as opposed to bound states.
But for suitably large angular momenta the half lives can be pushed up
as high as desired and the states will be extremely long lived.  Such
states are appropriate for a quantum description of a particle in a
classically stable orbit some distance from the horizon.

\begin{figure}
\begin{center}
\includegraphics[width=6cm]{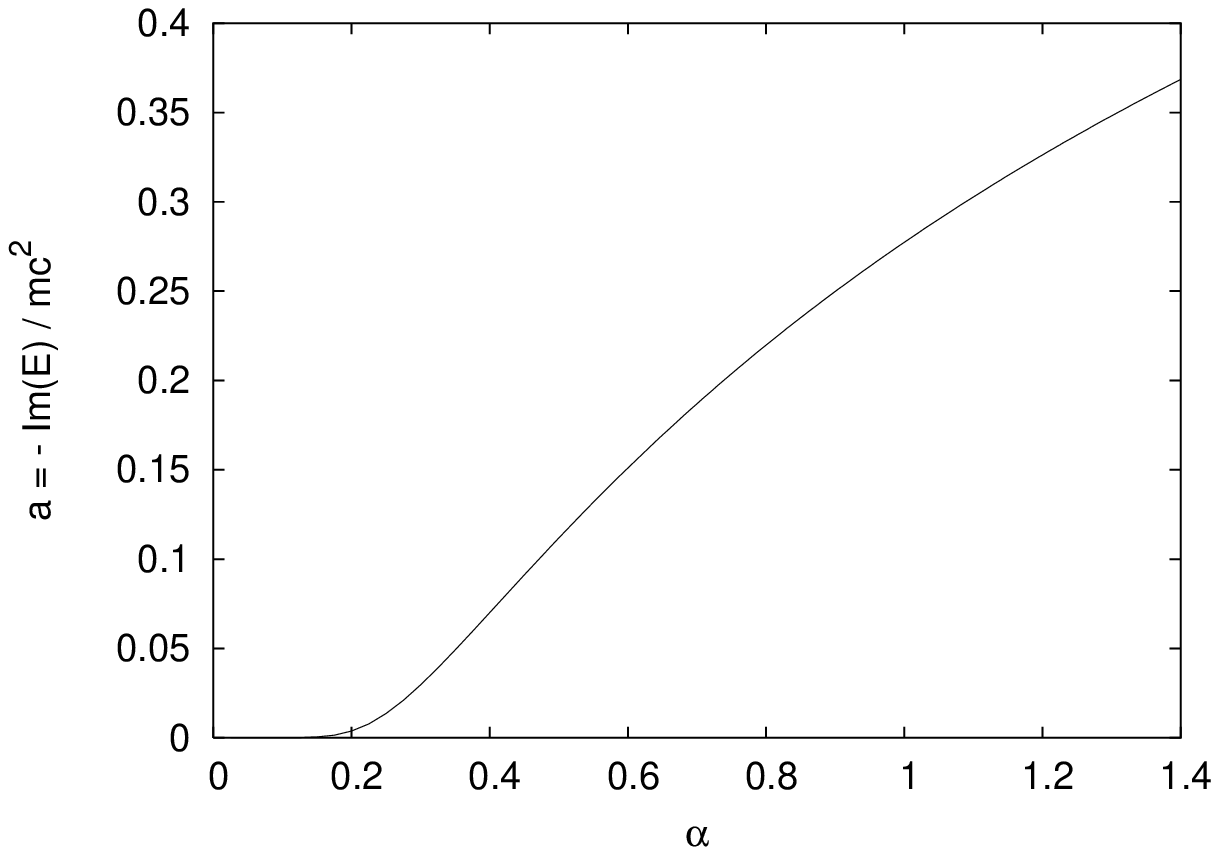}
\includegraphics[width=6cm]{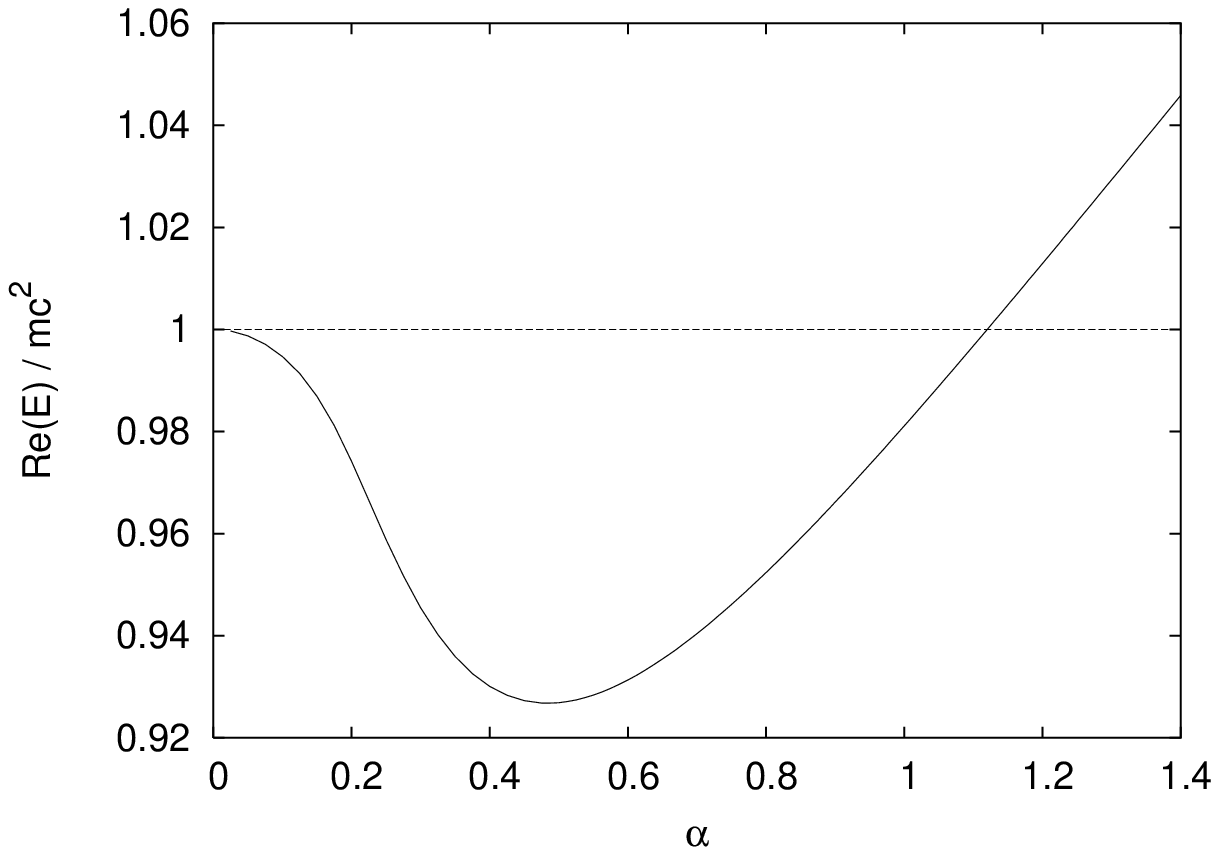}
\end{center}
\caption[dummy1]{\textit{The imaginary and real energies for the $1S_{1/2}$ state}.  
The left hand plot shows (minus) the imaginary component of the
energy as a function of the coupling strength.  As expected, this
increases as the orbits become more tightly bound.  For comparison,
the more complicated behaviour of the real part of the energy is shown
on the right-hand side.}
\label{FEi1}
\end{figure}

As argued above, the imaginary part of the energy will be negative,
corresponding to a decay.  The behaviour of this decay can be
visualised in a number of ways.  With $E=\om-i\nu$, the relevant
quantity to study is 
\begin{equation}
a = \frac{\nu}{mc^2} .
\end{equation}
In figure~\ref{FEi1} we plot $a$ as a function of coupling for the
$1S_{1/2}$ state.  For comparison the real part of the energy is also
plotted.  The real energy falls to a minimum and starts increasing
again as the orbits become unfavourably close, whereas the imaginary
term simply increases monotonically.  This as one would expect, as
figure~\ref{Fexpr} showed that the orbits become increasingly tightly
bound as $\alp$ increases.  As the coupling strength reaches $1$, the
imaginary component of the energy is of the order of $0.3$ times the
rest energy of the particle.  This implies that the orbit should decay
on the time-scale defined by the Compton frequency.  These states are
therefore extremely short lived, with a resonance width comparable to
the orbital energy.

\begin{figure}
\begin{center}
\includegraphics[width=10cm]{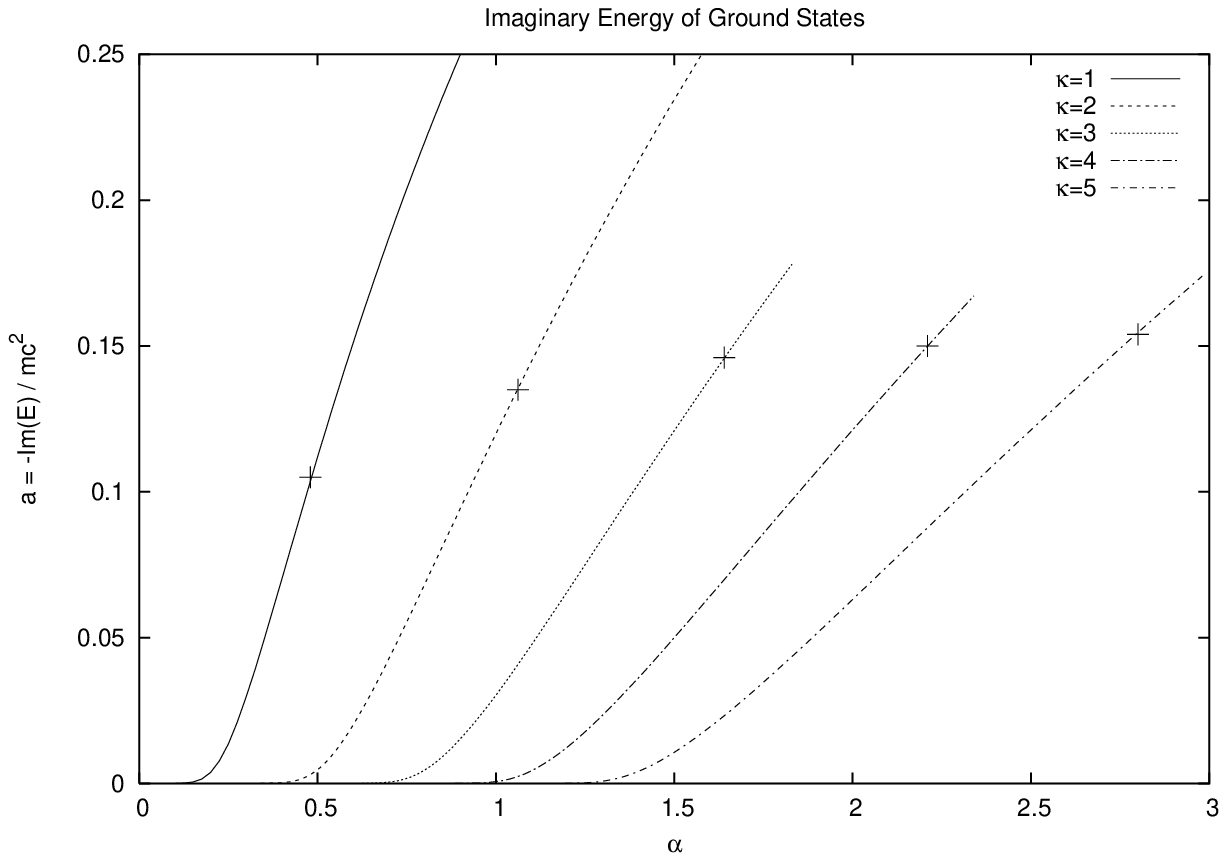}
\includegraphics[width=10cm]{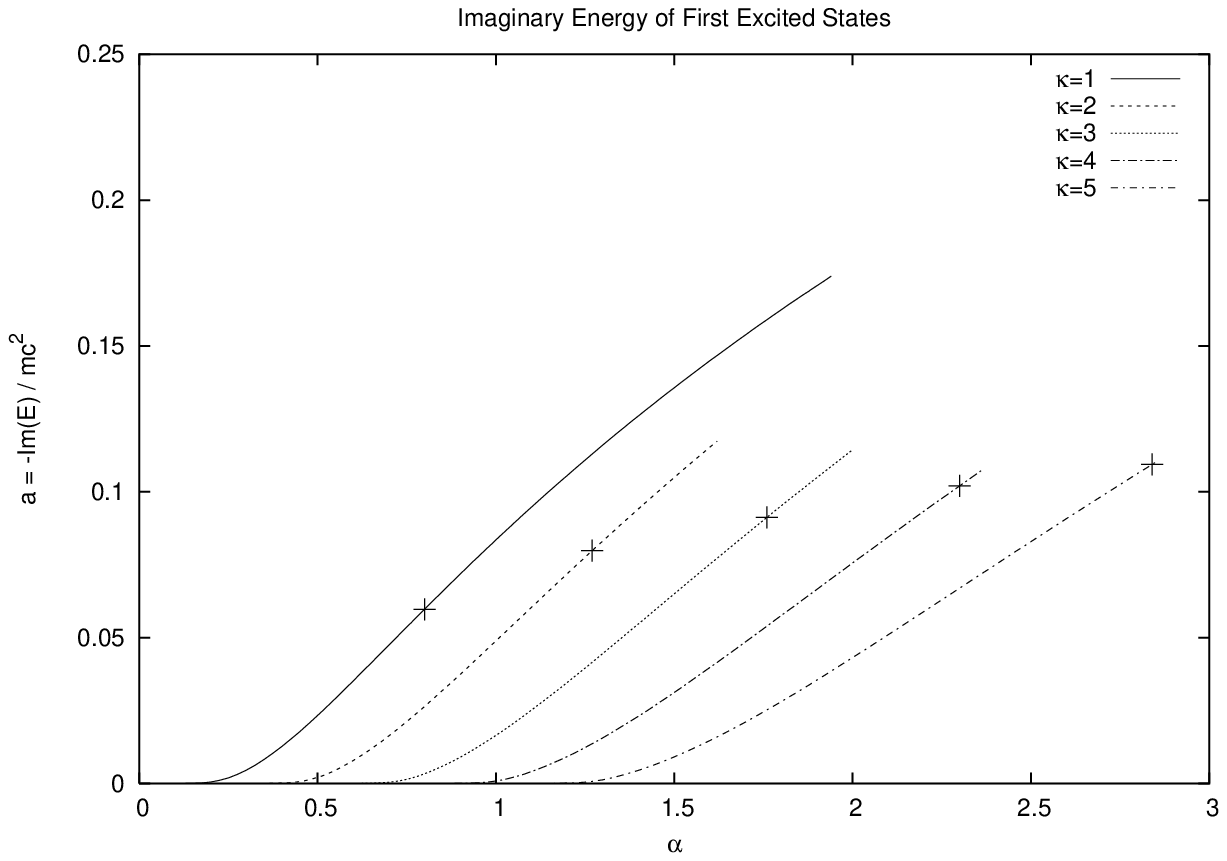}
\end{center}
\caption[dummy1]{\textit{The imaginary energies of states with a range of angular momenta $\kappa = 1 \cdots 5$}. 
The top plot shows the decay rates of the ground states, and the bottom plot shows
the decay rates of the first-excited states, as functions of the coupling strength $\alpha$. The positions of the minima in the real energy are marked with crosses.}
\label{FEi2}
\end{figure}

In figure~\ref{FEi2} $a$ is plotted for states with a range of angular
momenta, $\kappa = 1 \cdots 5$.  The set of lowest-energy states
($1S_{1/2}$, $2P_{3/2}$,\ldots) is compared to the set of
first-excited states ($2S_{1/2}$, $3P_{3/2}$,\ldots).  Both plots
show the expected monotonic increase in $a$ with coupling strength as
the orbits become more tightly bound and a greater percentage of the
wavefunction lies inside the horizon. The first-excited states are
less tightly bound than the ground states, so have smaller decay
rates.  Below a threshold value of $\alp$, the imaginary energy is
negligible.  This threshold depends roughly linearly on $\kappa$, and
is the same for the lowest and first-excited states.  Using the
effective-potential model, we would expect decay to become dominant
beyond the last value of $\alpha$ that allows stable circular orbits,
$\alpha = \kappa / \sqrt{12} = 0.29 \kappa$. The plot suggests that
this model is reasonably valid. States with higher angular momentum
can therefore be extremely stable, as the increase in $\kappa$ keeps
the bulk of density away from the singularity.

A classical argument can also be used to relate the high-$\alpha$
behaviour of the imaginary energy to the expectation value of
wavefunction radius, by considering the proper time for radial
infall. A massive particle starting at radius $r_i$ from rest would
take proper time
\begin{equation}
\tau_{\mathrm{infall}} = \sqrt{\frac{{r_i}^3}{8 G M}} \pi
\end{equation}
to reach the singularity. Conversely, the typical decay time for the wavefunction is 
\begin{equation}
\tau_{\mathrm{decay}} = \frac{\hbar}{a m c^2} 
\end{equation}
If the decay time is similar to the infall time from the wavefunction
expectation position $\langle x \rangle$, we would expect
\begin{equation}
a \alpha \propto {\langle x \rangle}^{-3/2}
\end{equation}
This model works well for the $1S_{1/2}$ state, and we find $a \alpha
\propto {\langle x \rangle}^{-1.6}$ in the high-$\alpha$ regime. The
model requires some modifications for states with orbital angular
momentum, as the infall time takes a more complicated form.

With the decay rates now obtained, we can return to
equation~\eqref{nufrmnorm} to check the consistency of our method.
For a number of states we computed the normalization integral and also
extracted the behaviour of the state near the singularity.  For all of
these the imaginary component of the energy was consistent with
equation~\eqref{nufrmnorm}.  This confirms that the states are
normalizable and represent genuine bound states.

\section{Discussion}
\label{disc}

We have demonstrated the existence of a complicated spectrum of bound
states for a quantum fermion in a black hole background.  Each state
represents a spatially-normalizable solution to the Dirac equation in
a Schwarzschild background.  The fact that time-separable solutions
exist is simply established in one particular gauge, which casts the
equation in a Hamiltonian-like form.  A study of the behaviour of the
wavefuntion under gauge transformations show that time-separability is
a gauge-invariant feature.  The spectrum itself is determined by
boundary conditions applied at the horizon and at infinity.  These
alone are sufficient to imply the existence of an imaginary (decay)
contribution to the energy.  The physical explanation for this is
provided by the singularity, which acts as a current sink.

The qualitative features of the spectrum can be understood in terms of
simple semi-classical models, but a full quantitative understanding
only seems possible through a mixture of computational methods.  The
work in this paper can clearly be extended in a number of ways.  We
have only plotted the spectrum at low coupling strengths of $\alpha
\sim 1$, but astrophysical values can be far larger than this, with
$\alpha \sim 10^{15}$ for solar mass black holes.  For larger
$\alpha$, the ground state will be one of high angular momentum.  In
this regime the spectrum will be quite different to that of the
hydrogen atom.  One important question is precisely how great a
binding energy can be achieved.  In figure~\ref{higher_kappa} we see
that at around $\alp=5$ we are achieving total energies of $0.88mc^2$,
which is significantly lower the classical value of $0.94mc^2$.  This
suggests that more energy may be available in accretion processes than
is traditionally thought.

As well as increasing $\alp$, it would be of considerable interest to
repeat this work for the case of a Kerr black hole.  In this respect a
useful start has been made in~\cite{D00-kerr}, where the Kerr solution
is written in a form which generalises the `Newtonian' gauge employed
in this paper.  The calculations for the Kerr case are more
complicated, however, because the angular separation constants are
energy-dependent~\cite{cha83,gai88}.  There are also signs that the
horizon structure of the Kerr solution will complicate the fairly
straightforward picture presented here.  The problem can be seen by
analysing behaviour in a Reissner--Nordstrom background using the
setup of this paper.  For this case we find that the regular solutions
at the outer horizon do not match onto regular solutions at the inner
horizon.  So quantum mechanics predicts that the probability density
will pile up around the inner horizon in a similar manner to the
classical picture.  Behaviour of this type is inevitable, as the
Reissner--Nordstrom singularity does not act as a sink, and the
Hamiltonian is Hermitian.  Since the current must still be inward
pointing at the outer horizon, the probability density has to pile up
somewhere.  It seems likely that a similar picture holds for the Kerr
solution, but detailed calculations are required to confirm this.

The energy spectra presented in this paper raise a number of
fundamental issues, which demonstrate the limitations in our current
understanding of the interaction between gravity and quantum theory.
It is unusual to obtain a decay law from quantum mechanics without
some form of approximation.  That we do so in the present case is a
consequence of the fact that the system is open.  States are allowed
to decay onto the singularity, but no accompanying emission is
considered.  A complete treatment of the problem as a closed
system would require a quantum theory of the singularity,  and
such a theory does not yet exist.

The decay rates represent one feature of the quantum-mechanical
description of the capture process.  But, as well as decay, the quantum
description of a particle falling onto the singularity of a black hole
can involve a series of quantum jumps to lower energy orbits.  This
quantum description alters the physics of the process quite
dramatically from the classical picture.  As the particle undergoes a
series of transitions we expect that it should radiate, which does not
happen classically.  Quite what form this radiation should take
(electromagnetic, gravity waves?) is unclear.  Also, as a transition
takes place we should keep careful track of the evolution of the
matter stress-energy tensor to tell us where the radiated energy is
concentrated.  A related problem this exposes is that we have not
considered back reaction on the gravitational field, which could alter
this picture.

The quantum treatment of a particle in a gravitational field exhibits
a curious anti-parallelism with the electromagnetic case.  In classical
electrodynamics a charged particle in orbit around a point source
should radiate, making atoms unstable.  This problem is resolved by
quantum mechanics, which predicts the existence of stable,
non-radiating bound states.  The reverse is true of gravitation.
Classically, a particle can orbit a black hole in a geodesic outside
the horizon, and such an orbit is stable.  But quantum theory changes
this, and states that no totally stable orbits exist, due to the
finite probability of the particle finding itself inside the horizon
and ending on the singularity.  While the time-scales involved in these
decays may be of limited interest astrophysically, such processes are
clearly of fundamental importance in understanding the interplay
between quantum theory and gravitation.

A final point to raise here is that the spectrum of real energies
derived here has a mirror image of negative energy bound states.  Each
of these negative energy states also has a finite lifetime.  If we
model the vacuum in terms of a Dirac sea of filled negative energy
states, we must include the bound states as well as the free
continuum.  It then follows that the vacuum itself is decaying --- the
black hole is sucking in the vacuum.  Such a loss of negative energy
states is seen as a creation of positive energy modes, which could
contribute to Hawking radiation.  This contribution appears to have
been neglected in previous calculations, which concentrate only on the
scattered states~\cite{bir-quant}.  It is well known in calculations
of the Lamb shift, for example, that ignoring the bound states in the
calculation gives the wrong answer~\cite{itz-quant}.  It would be of
great interest to assess the contribution played by bound states to
the gravitational analogue of this process.

\end{document}